\documentclass[prl,twocolumn,superscriptaddress,showpacs]{revtex4}
\usepackage{graphicx,amsmath,amssymb,bm}

\newcommand{\beqn}{\begin{equation}}
\newcommand{\eeqn}{\end{equation}}
\newcommand{\bea}{\begin{eqnarray}}
\newcommand{\eea}{\end{eqnarray}}
\newcommand{\ba}{\begin{align}}
\newcommand{\ea}{\end{align}}
\newcommand{\lm}{\Lambda}

\newcommand{\fm}{\, \text{fm}}
\newcommand{\fmi}{\, \text{fm}^{-1}}

\newcommand{\mev}{\, \text{MeV}}
\newcommand{\vlowk}{V_{{\rm low}\,k}}
\newcommand{\kf}{k_{\rm F}}

\begin{document}

\title{Nuclear matter from chiral low-momentum interactions}

\author{S.\ K.\ Bogner}
\email{bogner@nscl.msu.edu}
\affiliation{National Superconducting Cyclotron Laboratory and
Department of Physics and Astronomy, Michigan State University, 
East Lansing, MI 48844, USA}
\author{R.\ J.\ Furnstahl}
\email{furnstahl.1@osu.edu}
\affiliation{Department of Physics, The Ohio State University, 
Columbus, OH 43210, USA}
\author{A.\ Nogga}
\email{a.nogga@fz-juelich.de}
\affiliation{Institute for Advanced Simulations, 
Institut f\"ur Kernphysik and J\"ulich Centre for Hadron Physics,
Forschungszentrum J\"ulich, 52425 J\"ulich, Germany}
\author{A.\ Schwenk}
\email{schwenk@triumf.ca}
\affiliation{TRIUMF, 4004 Wesbrook Mall, Vancouver, BC, V6T 2A3, Canada}

%\date{\today}

\begin{abstract}
Nuclear matter calculations based on low-momentum
interactions derived from chiral nucleon-nucleon
and three-nucleon effective field theory
interactions and fit only to few-body data predict
realistic saturation properties with
controlled uncertainties. This is promising
for a unified description of nuclei and to develop
a universal density functional based on
low-momentum interactions.
\end{abstract}


\maketitle

Nuclear forces saturate, so nuclei are self bound with roughly constant
interior density.
The Coulomb interaction drives heavier stable nuclei toward an
imbalance of
neutrons over protons and eventual instability, but 
theorists can extrapolate a Coulomb-free, $N=Z$ nucleus to 
arbitrary size.
The uniform limit is called symmetric nuclear matter.
For fifty years, an accurate prediction of nuclear matter starting
from nuclear forces has been 
a theoretical milestone on the way to finite nuclei,
but has proved to be an elusive target.
Here we present the first nuclear matter calculations using 
soft Hamiltonians derived from chiral effective field theory
interactions fit only to few-body ($A \leqslant 4$) data.
We find realistic saturation properties with
controlled uncertainties. 

Despite the long-term emphasis on the infinite uniform system,
most advances in microscopic nuclear structure theory over the
last decade have been through 
expanding the reach of few-body calculations.
This has unambiguously established the quantitative
role of three-nucleon forces (3NF) for light nuclei 
($A \leqslant 12$)~\cite{GFMC,NCSMchiral}.
However, until now few-body fits
have not sufficiently constrained 3NF contributions at
higher density such that nuclear matter calculations are predictive. 
Nuclear matter saturation is very delicate, with the binding energy
resulting from
cancellations of much larger potential and kinetic energy contributions.
When a quantitative reproduction of empirical saturation properties
has been obtained, it was
imposed by hand through adjusting
short-range three-body forces (see, for example, 
Refs.~\cite{Akmal:1998cf,Lejeunenm}).
%~\cite{Pudliner:1997ck,Akmal:1998cf}

Progress for controlled nuclear matter calculations
has long been hindered by the difficulty
of the nuclear many-body problem when conventional
nuclear potentials are used.
The present calculations pull together several recent developments
to overcome the hurdles:
systematic starting Hamiltonians based on chiral
effective field theory (EFT)~\cite{N3LO,N3LOEGM}, 
renormalization group (RG) methods~\cite{Vlowk,smooth} to soften
the short-range repulsion and short-range tensor components of the
initial chiral interactions~\cite{Born} so that convergence of 
many-body calculations is vastly accelerated~\cite{nucmatt,NCSM,Sonia},
and an improved 3NF fitting procedure to the $^4$He 
radius~\cite{NCSMchiral}. In combination we obtain controlled
theoretical uncertainties.

\begin{figure*}[t]
\begin{center}
\includegraphics[scale=0.44,clip=]{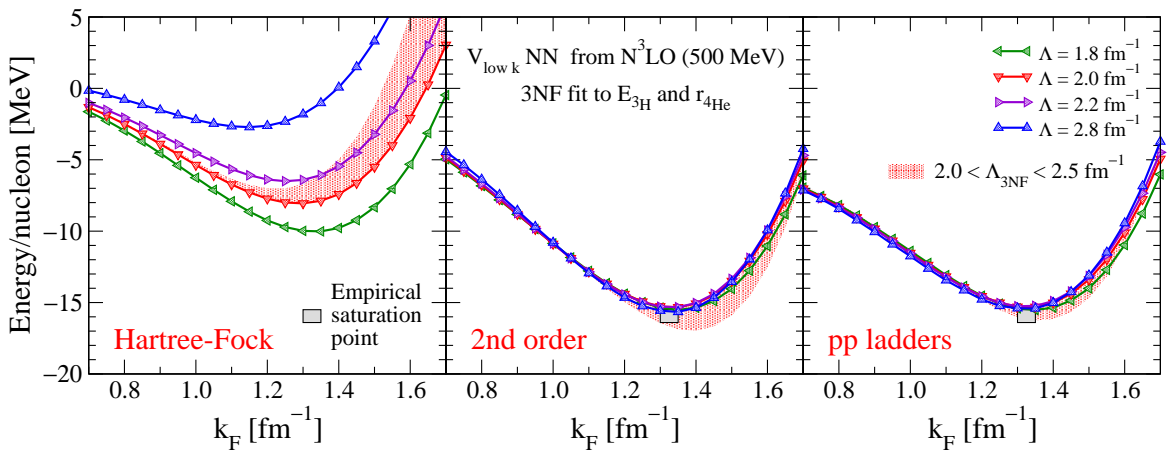}
\end{center}
\caption{(Color online) Nuclear matter energy per particle as a
function of Fermi momentum $\kf$ at the Hartree-Fock level (left)
and including second-order (middle) and particle-particle-ladder 
contributions (right), based on evolved N$^3$LO NN potentials and
3NF fit to $E_{\rm^3H}$ and $r_{\rm^4He}$. Theoretical 
uncertainties are estimated by the NN (lines) and 3N (band)
cutoff variations.\label{nm_all}}
\end{figure*}

Our results are summarized in Fig.~\ref{nm_all},
which shows the energy per particle of symmetric matter
as a function of Fermi momentum $\kf$, or the density $\rho 
= 2 \kf^3/(3\pi^2)$.
A grey square representing the empirical saturation 
point is shown in each of the nuclear matter figures.
Its boundaries reflect the ranges of nuclear matter saturation
properties predicted by phenomenological Skyrme energy functionals that
most accurately reproduce properties of finite nuclei.
Although this determination cannot be completely model independent, 
the value is generally accepted for benchmarking infinite matter.
In the future, calculations of the properties of finite nuclei 
will allow one to compare directly to experimental data.

The calculations of Fig.~\ref{nm_all} start from the
N$^3$LO nucleon-nucleon (NN) potential (EM $500 \mev$) of 
Ref.~\cite{N3LO}. This NN potential is RG-evolved to
low-momentum interactions $\vlowk$ with a smooth 
$n_{\rm exp}=4$ regulator~\cite{smooth}. For each 
cutoff $\Lambda$, two couplings that determine the 
shorter-range parts of the ${\rm N^2LO}$ 
3NF~\cite{chiral3N} %, $c_D$ and $c_E$ of Eq.~(\ref{3NF}),
are fit to the $^3$H binding energy and the $^4$He matter
radius using exact Faddeev and Faddeev-Yakubovsky
methods as in Ref.~\cite{Vlowk3N}. Our 3NF fit
values are given in
Table~\ref{3Nfits}. We use the same 3NF regulator
$\exp[-(p^2 + 3/4 q^2)^2/\lm_{\rm 3NF}^4]$ of
Ref.~\cite{chiral3N}, but with a 3N cutoff $\lm_{\rm 3NF}$
that is allowed to vary independently of the NN cutoff.
The shaded regions in Fig.~\ref{nm_all} show
the range of results for $2.0 \fmi < \Lambda_{\rm 3NF} < 
2.5 \fmi$ at fixed $\Lambda = 2.0 \fmi$. Nuclear matter
is calculated in three approximations:
Hartree-Fock (left) and including
approximate second-order (middle) and summing 
particle-particle-ladder contributions (right). 
The technical details are given in Ref.~\cite{nucmatt},
but we have improved the calculation to include full 
momentum-dependent Hartree-Fock propagators and
(sub-leading) 3N double-exchange diagrams beyond 
Hartree-Fock, however in a very approximate way.
%with a scale factor matched to the exact 
%Hartree-Fock energy
Further improvements are in progress~\cite{3Nnm}.
%and more details
%will be given in a longer paper

The Hartree-Fock results show that nuclear matter is bound even at
this simplest level.
A calculation without approximations should be independent of the
cutoffs, so the spread in Fig.~\ref{nm_all}
sets the scale for omitted many-body
contributions. The second-order results show a dramatic 
narrowing of this spread, with predicted saturation 
consistent with the empirical range. 
The narrowing happens across the full density range. This 
is strong evidence that these encouraging results are not 
fortuitous, but that we have reached cutoff independence 
at the level of $1-2 \mev$ per particle.
The 3NF fits to the $^4$He radius improve the cutoff
independence significantly compared to fitting to $A=3,4$
energies only, see Fig.~6 in Ref.~\cite{nucmatt}.
For all cases, the compressibility $K = 190-240 \mev$
(mainly from $\Lambda_{\rm 3NF} = 2.0-2.5 \fmi$)
is in the empirical range.
To our knowledge, these are the first
nuclear forces fit only to $A \leqslant 4$ nuclei that 
predict saturation reasonably.

The particle-particle-ladder sum is little changed from 
second order except at the lowest densities shown. This
is not surprising because at low density the presence of 
a two-body bound state necessitates a nonperturbative summation.
We note that below saturation density, the ground state of
nuclear matter is not a uniform system, but breaks into 
clusters (see, for example, Ref.~\cite{virial}).
 
In chiral EFT without explicit Deltas, 3N interactions
start at N$^2$LO~\cite{chiral3N} and their contributions
are given diagrammatically by
\beqn
\parbox[c]{195pt}{%
\includegraphics[scale=0.6,clip=]{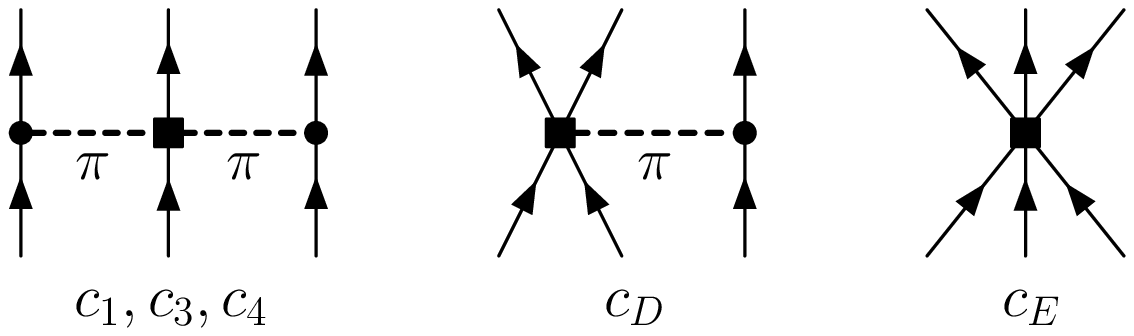}}
\nonumber
%\label{3NF}
\eeqn
We assume that the $c_i$ coefficients of the long-range
two-pion-exchange part are not modified by the RG and
take these values from Ref.~\cite{Rentmeester:2003mf}.  
At present, we rely on the N$^2$LO 3NF as a truncated
``basis'' for low-momentum 3N interactions and
determine the $c_D$ and $c_E$ couplings by a fit to data
for all cutoffs~\cite{Vlowk3N}. In the future, it will be
possible to fully evolve three- and four-body forces 
starting from chiral EFT~\cite{SRG3b,1d}. The fit values
of Table~\ref{3Nfits} are natural and the predicted $^4$He
binding energies are very reasonable. We have also verified
that the resulting 3NF becomes perturbative in $A=3,4$
(see also Ref.~\cite{Vlowk3N}) for the cutoffs used 
and generally for $\lm_{\rm 3NF} \lesssim 
\lm$~\cite{3Nnm}. The sensitivity of many-body observables
to uncertainties in the $c_i$ coefficients was
demonstrated recently in neutron matter 
calculations~\cite{nm}. This raises the
possibility of using nuclear matter to constrain some 
of the $c_i$ couplings.

\begin{table}[t]
\begin{ruledtabular}
\begin{tabular}{c|p{1.225cm}p{1.225cm}|p{1.225cm}p{1.225cm}}
& \multicolumn{2}{c|}{$\vlowk$} & \multicolumn{2}{c}{SRG} \\[0.2mm] \hline
$\Lambda$ or $\lambda/\lm_{\rm 3NF}$ [fm$^{-1}$]
& \multicolumn{1}{c}{$c_D$} & \multicolumn{1}{c|}{$c_E$} 
& \multicolumn{1}{c}{$c_D$} & \multicolumn{1}{c}{$c_E$} \\[0.2mm] \hline
$1.8/2.0$ & $-0.0112$ & $-0.2212$ & & \\
$2.0/2.0$ & $-0.3000$ & $-0.2761$ & $-1.023$ & $-0.3397$ \\
$2.0/2.5$ & $-2.000$ & $-0.7564$ & $-2.991$ & $-0.8797$ \\
$2.2/2.0$ & $-0.9000$ & $-0.3673$ & & \\
$2.8/2.0$ & $-1.552$ & $-0.4058$ & & \\
\end{tabular}
\end{ruledtabular}
\caption{Results for the $c_D$ and $c_E$ couplings fit to $E_{^3{\rm H}} 
= -8.482 \mev$
and $r_{^4{\rm He}} = 1.95-1.96 \fm$ for the NN/3N cutoffs used here.
For $\vlowk$ (SRG) interactions, the 3NF fits lead to $E_{^4{\rm He}} = 
-28.22\ldots28.45 \mev$ ($-28.53\ldots28.71 \mev$).\label{3Nfits}}
\end{table}

\begin{figure}[t]
\begin{center}
\includegraphics[scale=0.37,clip=]{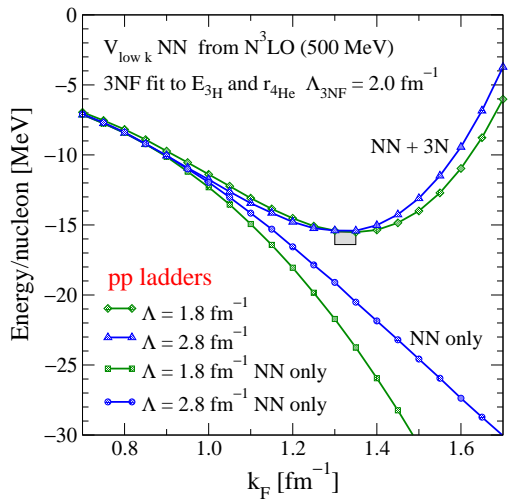}
\end{center}
\caption{(Color online) Nuclear matter energy of 
Fig.~\ref{nm_all} at the particle-particle-ladders level compared
to NN-only results for two representative NN cutoffs and a fixed
3N cutoff.\label{NNvs3N}}
\end{figure}

The evolution of the cutoff $\Lambda$ to smaller values is
accompanied by a shift of physics. In particular, effects due
to iterated tensor interactions, which peak in the relative
momentum range $k \sim 4 \fmi$ (and thus lead to saturation at
too high density), are replaced by three-body contributions.
The role of the 3NF for saturation is demonstrated in
Fig.~\ref{NNvs3N}. The two pairs of curves show the difference
between the nuclear matter results for NN-only and NN plus 3N
interactions. It is evident that saturation is driven by the
3NF~\cite{nucmatt}. Even for $\Lambda = 2.8 \fmi$, which is 
similar to the lower cutoffs in chiral EFT 
potentials, saturation is at too high density without 
the 3NF. Nevertheless, as in previous results~\cite{nucmatt},
the 3N contributions and the $c_D, c_E$ fits are natural,
and the same is expected for the ratio of three- to four-body 
contributions.
 
\begin{figure}[t]
\begin{center}
\includegraphics[scale=0.37,clip=]{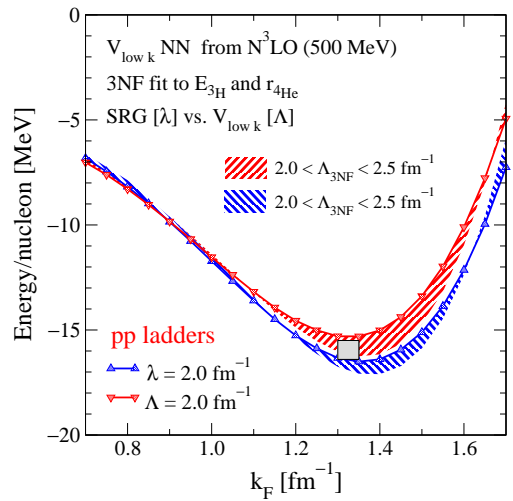}
\end{center}
\caption{(Color online) Nuclear matter energy of 
Fig.~\ref{nm_all} at the particle-particle-ladders level 
comparing low-momentum $\vlowk$ with SRG-evolved chiral
NN interactions.\label{nmsrg}}
\end{figure}

The smooth RG evolution used in the results
so far is not the only choice for low-momentum interactions.
A recent development is the use of flow equations to evolve
Hamiltonians, which we refer to as the Similarity Renormalization
Group (SRG)~\cite{SRG,SRGnuc,Roth:2005pd}.
The flow parameter $\lambda$, which has dimensions of a momentum,
is a measure of the degree of decoupling in momentum space.
Few-body results for roughly the same value of SRG $\lambda$ and
smooth $\vlowk$ $\Lambda$ have been remarkably similar (see,
for example, Ref.~\cite{NCSM}).
The analogous nuclear matter energies shown in
Fig.~\ref{nmsrg} are also similar, which helps support the 
general nature of the 3NF fit.  On the other hand, the difference
of $2 \mev$ per particle at saturation and above enlarges our
theoretical uncertainty.
 
\begin{figure}[t]
\begin{center}
\includegraphics[scale=0.4,clip=]{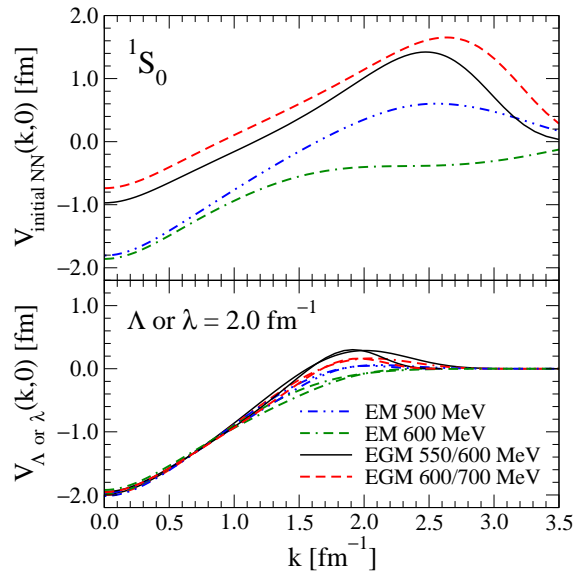}
\end{center}
\caption{(Color online) Off-diagonal momentum-space matrix
elements of different chiral N$^3$LO potentials (EM~\cite{N3LO}
and EGM~\cite{N3LOEGM}) in the $^1$S$_0$ channel (upper panel)
and after evolution to low-momentum $\vlowk$ or SRG interactions
(lower panel). A similar universal collapse is found for the
low-momentum
diagonal matrix elements and in other channels.\label{collapse}}
\end{figure}

The presented results are starting from a chiral EFT potential 
at a single
order with one choice of EFT regulator implementation and values.
There are several alternatives available~\cite{N3LO,N3LOEGM,chiral},
which are largely phase-shift equivalent (the chi-square is not
equally good up to $E_{\rm lab} \approx 300 \mev$).
Universality for phase-shift equivalent chiral EFT potentials
was shown for smooth-cutoff $\vlowk$ interactions in 
Ref.~\cite{smooth} in the form of a collapse of the different 
potentials to the same matrix elements. An analogous collapse 
for N$^3$LO potentials evolved by the SRG is shown in 
Fig.~\ref{collapse} for the $^1$S$_0$ channel, with a
comparison to the corresponding $\vlowk$ interactions.
Similar results are found in the other channels.
Based on this universal collapse,
we do not anticipate large differences in nuclear matter when
starting the evolution with different N$^3$LO potentials, 
but a detailed comparison is forthcoming~\cite{3Nnm}.
 
The theoretical errors of our results
arise from truncations in the initial chiral EFT Hamiltonian,
the approximation of the 3NF, and the many-body approximations.
We do not claim a chiral expansion for nuclear matter saturation,
only that the hierarchy of potential energy contributions is
maintained by the RG/SRG evolution.
Is the nuclear matter many-body calculation under control?
Corrections to the present approximation include higher-order terms
in the hole-line expansion and particle-hole corrections.
While we have positive circumstantial evidence from
cutoff independence that these corrections are small,
we believe that an approach such as coupled cluster theory that can
perform a high-level resummation is necessary for a robust
validation.

While nuclear matter has lost its status to light nuclei as the first step
to nuclear structure,
it is still key as a step to heavier nuclei.
Our results open the door
to ab-initio density functional theory (DFT)
based on expanding about nuclear matter~\cite{DME}.
This is analogous to the application of
DFT in quantum chemistry and condensed matter starting with the uniform
electron gas in local-density approximations and adding constrained derivative
corrections.  Phenomenological energy functionals (such as Skyrme) 
for nuclei have
impressive successes but lack a (quantitative) microscopic foundation
based on
nuclear forces and seem to have reached the limits of
improvement with the current form of 
functionals~\cite{Bertsch:2004us,Kortelainen:2008rp}.
The theoretical errors of our results, while impressively
small on the scale of the potential energy per particle,
are far too large to be quantitatively competitive with 
existing functionals. However, there is the possibility of fine 
tuning to heavy nuclei, of using EFT/RG to guide next-generation
functional forms, and of benchmarking with ab-initio methods 
for low-momentum interactions. Work in these directions is in progress.

In summary, we have presented the first results for nuclear matter
based on chiral NN and 3N interactions with RG evolution.
The chiral EFT framework provides a systematic improvable Hamiltonian
while the softening of nuclear forces by RG evolution
enhances the convergence and control of the many-body calculation.
The empirical saturation point is reproduced with theoretical
uncertainties despite input only from few-body data.
Because of the fine cancellations, however, significant reduction of
the errors will be needed before direct DFT calculations of nuclei
are competitive.
Nevertheless, these results are very promising for a unified
description of all nuclei and nuclear matter.  

\begin{acknowledgments}
%We thank ??? for useful comments.
This work was supported in part by the National Science Foundation
under Grant Nos.~PHY--0354916 and PHY--0653312, the UNEDF SciDAC 
Collaboration (unedf.org) under DOE Grant DE-FC02-07ER41457, and the Natural 
Sciences and Engineering Research Council of Canada (NSERC). TRIUMF
receives federal funding via a contribution agreement through the 
National Research Council of Canada. Part of the numerical calculations
have been performed at the JSC, J\"ulich, Germany.
\end{acknowledgments}

\end{document}